\documentclass[12pt]{iopart}
\usepackage{graphicx}

\begin{document}

\title[Boronization-enabled I-mode on EAST tokamak with an expanded density window and favorable-configuration access]{Boronization-enabled I-mode on EAST tokamak with an expanded density window and favorable-configuration access}

\author{X.M.~Zhong$^{1}$, X.L.~Zou$^{2}$, A.D.~Liu$^{3,*}$, L.Q.~Xu$^{1,*}$, B.~Zhang$^{1}$, C.~Zhou$^{3}$, J.P.~Qian$^{1}$, X.Z. Gong$^{1}$, Y.T.~Song$^{1}$, G.~Zhuang$^{3}$, W.X.~Shi$^{3}$, L.T.~Gao$^{3}$, S.~F.~Wang$^{3}$, Y.H. Guan$^{1}$, G.Z. Zuo$^{1}$, T.Q.~Jia$^{1}$, Y.X.~Cheng$^{1}$, S.X.~Wang$^{1}$, K.N.~Geng$^{1}$, H.L.~Zhao$^{1}$, EAST I-mode Working Group$^{\rm a}$, and EAST Team$^{\rm b}$}

\address{$^{1}$ Institute of Plasma Physics, Chinese Academy of Sciences, Hefei 230031, China}
\address{$^{2}$ CEA, IRFM, F-13108 Saint-Paul-lez-Durance, France}
\address{$^{3}$ School of Nuclear Science and Technology, University of Science and Technology of China, Hefei 230026, China}

\ead{lad@ustc.edu.cn, lqxu@ipp.ac.cn}


\begin{abstract}
I-mode is a promising confinement regime for future fusion reactors because it combines enhanced energy confinement with L-mode-like particle transport and naturally ELM-free operation. 
Previous EAST I-mode studies were performed exclusively under lithium-conditioned wall conditions. 
Here we report the first systematic experimental investigation of I-mode under boronized wall conditions on EAST and compare it with an existing lithium-conditioned I-mode database at the same toroidal field, $B_t = 2.47$\,T. 
The boronized-wall dataset exhibits a substantially broader accessible density range, with the Greenwald fraction extending from  $f_{\mathrm{GW}} = 0.26 - 0.77$ , compared with $f_{\mathrm{GW}} = 0.35 - 0.54$ under lithiation.  
A higher normalized $\mathrm{D}_\alpha$ emission suggests that enhanced edge recycling may contribute to this density extension. A striking increase in favorable-configuration I-mode is also observed: $51\%$ boronized-wall discharges are obtained in favorable-configuration, compared with only $8\%$ lithium-conditioned discharges. 
These favorable-configuration cases are concentrated at high density and exhibit a deeper radial electric-field($E_r$) well and stronger $\mathbf{E_r}\times\mathbf{B}$ velocity shear. When ETRO is present, the associated transition between electron and ion turbulence is similar under the two wall conditions, although ETRO occurs less frequently ($15\%$) under boronization. 
An empirical EAST I-mode energy confinement scaling at fixed $B_t$ is obtained, $\tau_E = 3.29 I_p^{0.51 \pm 0.10} P_{\mathrm{loss}}^{-0.53 \pm 0.05} \bar{n}_e^{0.08 \pm 0.07}$, indicating weaker power degradation than IPB98(y,2) H-mode scaling and a weak density dependence. 
These results show that boronization can broaden the operational space of EAST I-mode and support the development of reactor-relevant ELM-free scenarios.

\end{abstract}

\vspace{2pc}

\submitto{\NF}
\par\medskip
\noindent$^*$ Authors to whom any correspondence should be addressed.\\
$^{\rm a}$ See Liu et~al 2024 ({\it Nucl. Fusion} {\bf 64} 016013) for the EAST I-mode Working Group.\\
$^{\rm b}$ See Gong et~al 2024 ({\it Nucl. Fusion} {\bf 64} 126019) for the EAST Team.


\section{Introduction}

The earliest observations of I-mode-like plasmas were reported on Alcator C-Mod~\cite{Greenwald1997NF} and ASDEX Upgrade (AUG)~\cite{Ryter1998PPCF}, where improved energy confinement with a temperature pedestal but no density pedestal was noted. The regime was subsequently named ``I-mode'' (improved energy confinement mode). It was formally defined as a plasma regime in which an edge energy transport barrier exists without an accompanying particle transport barrier, resulting in an H-mode-like temperature pedestal and an L-mode-like density profile~\cite{Whyte2010NF}. This natural decoupling of particle and heat transport produces a regime that is inherently ELM-free, as the pedestal pressure remains below the peeling-ballooning stability boundary~\cite{Walk2014POP}. Owing to the absence of a density pedestal, impurities are efficiently transported out of the core, thereby preventing core impurity accumulation~\cite{Whyte2010NF}. Such effective avoidance of impurity accumulation, together with the absence of ELMs, makes I-mode a compelling candidate for future burning plasma devices. Since its initial characterization, I-mode has been confirmed on multiple devices including ASDEX Upgrade (AUG)~\cite{Happel2019NME, Ryter2017NF}, Alcator C-Mod~\cite{Hubbard2011POP}, DIII-D~\cite{Marinoni2015NF}, EAST~\cite{Feng2019NF}, HL-2A~\cite{Liang2023NF}, and KSTAR~\cite{Chweeho2025NF}, establishing it as a robust operational scenario across a range of machine sizes and heating methods.

I-mode is typically accessed in the unfavorable-configuration, where $\mathbf{B}\times\nabla B$ ion drift direction pointing away from the active X point. Because a higher L--H transition power threshold is present in the unfavorable-configuration, a power window is created in which I-mode can be sustained. This regime is accompanied by broadband turbulence in both density and magnetic fluctuations, known as the weakly coherent mode (WCM), which is thought to provide a channel for enhanced particle transport~\cite{Whyte2010NF}. The WCM appears at frequencies of 40--300 kHz (device-dependent) in the pedestal region, and its presence in the edge fluctuation spectrum is widely used as a diagnostic signature confirming I-mode access. In addition, favorable-configuration I-mode was first reported on Alcator C-Mod~\cite{Hubbard2011POP}, but its operating window was narrower and the L--I transition power was lower than in the unfavorable-configuration. The radial electric field in both configurations was dominated by the diamagnetic contribution, with $E_r$ shear at an intermediate level between L-mode and H-mode. Furthermore, the I-mode operating space was extended into the high toroidal field range ($B_t$ up to 8 T) on Alcator C-Mod, where the power window between the L--I and I--H transitions was found to widen with $B_t$~\cite{Hubbard2017NF}. At $B_t > 7$ T, the I--H transition could not be triggered even with the full available heating power ($\sim$5 MW), indicating that the high magnetic field of future fusion devices is inherently favorable for I-mode access and sustainment. Additionally, a detailed comparison of L--I and I--H transition properties has been performed on AUG~\cite{Ryter2017NF}. The results indicate that the L--I power threshold in the unfavorable-configuration is approximately twice the L--H threshold in the favorable-configuration. This difference underscores the importance of investigating favorable-configuration I-mode. Moreover, an I-mode energy confinement scaling was obtained from dedicated dimensionless parameter scans on C-Mod~\cite{Wilks2019}. In this scaling, the power degradation exponent ($P^{-0.29}$) is found to be substantially weaker than that of H-mode ($P^{-0.69}$ in IPB98(y,2)~\cite{ITER1999NF}), which indicates that I-mode confinement degrades far more slowly with increasing heating power --- a property that is highly advantageous for reactor operation. Regarding density control, pellet-fueled I-mode plasmas have been demonstrated on AUG~\cite{Silvagni2023}. By pellet injection, the I-mode density was raised by 40--60\%, extending $f_{\mathrm{GW}}$ from 0.25--0.45 to 0.55--0.60, and a core Greenwald fraction of up to 0.8 was achieved. ELM-free operation and a confinement level of $H_{98}\sim 1.1$ were maintained throughout these discharges. This result provides a viable path toward I-mode operation at reactor-relevant densities.

Significant progress has also been made in I-mode studies on the EAST tokamak. First, I-mode was identified on EAST in 2019~\cite{Feng2019NF} and has since been extensively characterized by the EAST I-mode working group~\cite{ADLiu2024NF}. A systematic analysis of 65 I-mode discharges revealed that the L--I power threshold is slightly higher than the ITPA L--H scaling prediction and that the power degradation is weaker than in H-mode~\cite{YJLiu2020NF}. Moreover, as a key heating method, lower hybrid current drive (LHCD) was systematically investigated in EAST I-mode plasmas, demonstrating its capability for current profile control during I-mode operation~\cite{Wu2026NF}. Additionally, at the pedestal top, a novel edge temperature ring oscillation (ETRO, frequency 6--12\,kHz) was discovered and subsequently characterized in detail: it sustains steady-state I-mode by modulating quasi-periodic transitions between electron-diamagnetic-direction and ion-diamagnetic-direction turbulence~\cite{ADLiu2020NF}. Its $m=0$, $n=0$ structure, confirmed by multi-diagnostic measurements, distinguishes it from geodesic acoustic modes (GAM)~\cite{ADLiu2024NF}. Notably, compatibility between a core electron transport barrier and edge I‑mode has been observed on EAST, which is referred to as Super I‑mode~\cite{Song2023SciAdv}. Furthermore, I‑mode discharges have also been realized in helium plasmas, and the confinement is found to degrade with increasing helium concentration~\cite{ZhangB2021NF}. In addition, pedestal burst instabilities (PBI) were identified as density-gradient-driven precursors to the I--H transition, with a trigger threshold of normalized pedestal density gradient~\cite{Zhong2022NF}. Moreover, in the scrape-off layer, the statistical properties of blobs in EAST I-mode were found to be distinct from those in L-mode, reflecting the unique edge transport characteristics of the I-mode regime~\cite{Wang2023PST}. Furthermore, stable I-mode operation compatible with partial detachment has been demonstrated on EAST with neon seeding, extending I-mode toward reactor-relevant divertor conditions~\cite{Yu2025NF}. Under these conditions, core confinement is found to be maintained or even improved. This improvement is attributed to the cooling of the plasma edge and the resulting increase in the edge temperature gradient~\cite{Wu2026PPCF}. Moreover, real-time lithium powder injection was found to improve I-mode confinement through enhanced velocity shear and suppressed edge turbulence. A systematic classification of EAST I-mode into four types (Type~I--IV) was further established based on pedestal turbulence characteristics~\cite{Zhong2024NF}. 

A crucial caveat is that all of the above EAST results were obtained under lithium coating. Although lithium conditioning can effectively control impurity radiation and reduce the edge recycling level, its tritium retention rate is excessively high. On the other hand, glow discharge boronization has been adopted as the basic wall conditioning method in the new ITER baseline design for the need to effectively suppress tungsten sputtering and reduce radiative power loss~\cite{Loarte2025PPCF}. A systematic investigation of I-mode under boron coating, with direct comparison against the lithium coating, is therefore essential for evaluating the compatibility of I-mode with the material choices of next-step devices. In this paper we present the first systematic study of I-mode under boronization on EAST. We compare 37 I-mode discharges under boronization with 48 I-mode discharges under lithiation at matched toroidal field $B_t = 2.47$\,T.

This paper is organized as follows. Section 2 describes the experimental setup for I-mode under boronization on EAST. Section 3 presents a typical discharge, followed by a statistical I-mode comparison between boronization and lithiation in section 4. Section 5 examines the favorable-configuration I-mode. Section 6 reports the empirical confinement scaling. Discussion is shown in section 7, and conclusions are summarized in the last section.


\section{Experimental Setup}

Since 2020, EAST, China's first fully superconducting tokamak (major radius $R = 1.85$\,m, minor radius $a = 0.45$\,m), has been upgraded with an almost full metal wall~\cite{BGWang2023NME}. In this I-mode study, two coating methods were employed on EAST, namely, lithiation and boronization. Lithium coating was performed by vacuum evaporation of lithium from ovens mounted on a movable manipulator system~\cite{Zuo2013JNM}. During lithium deposition, the ovens were inserted to a position near the limiter; after coating, they were retracted. A typical coating session evaporated 10--30\,g of lithium over 1--2 hours, and was carried out before each experimental day. Moreover, boronization was performed using carborane ($\mathrm{C_2B_{10}H_{12}}$) as the boron source, with 10\,g of carborane consumed per boronization event~\cite{Liu2025NF,Guan2025NF}. The process was assisted by an ion cyclotron range of frequencies (ICRF) plasma. Following a single boronization, the boron-based coating exhibited a lifetime of approximately $10^{4}$\,s under plasma exposure, with a residual film thickness of approximately 30\,nm remaining post-campaign~\cite{Liu2025NF}.

The I-mode experiments with boron coating were conducted in the lower single null (LSN) configuration, preferentially chosen during this campaign because the upper divertor water-cooling system was not operational. During the boronization I-mode campaign, the dominant auxiliary heating methods were electron cyclotron resonance heating (ECRH), neutral beam injection (NBI), and lower hybrid wave (LHW) heating. The density was measured by a three-channel hydrogen cyanide (HCN) laser interferometer~\cite{ShenJ2013FED}, an 11-channel polarimeter-interferometer (POINT)~\cite{Liu2014RSI}, and a multi-channel density profile reflectometer~\cite{ZhangJ2023JINST}. The electron temperature $T_e$ profile was measured by a 32-channel ECE radiometer~\cite{LiuY2019EPJ}. The boron (B) radiation was measured by an extreme ultraviolet (EUV) spectrometer~\cite{Cheng2025JINST}. The absolute intensity of the $\mathrm{D}_\alpha$ line emission can be measured locally by the Filterscope system~\cite{XuZ2016RSI}.

The turbulence rotation and intensity were measured by an E-band multi-channel DR~\cite{Shi2025PPCF}. In addition, the perpendicular velocity, which is the sum of the $\mathbf{E}_r \times \mathbf{B}$ velocity ($V_{\mathbf{E}_r\times\mathbf{B}}$) to the lab frame and the phase velocity ($V_{\mathrm{phase}}$) of the density fluctuation, could be calculated by $V_{\perp} = 2\pi f_d / k_{\perp}$~\cite{Conway2004PPCF,Zou1999TR}, where $k_{\perp}$ is the wavenumber of the fluctuation, and $f_d$ is the Doppler shift. For some cases in~\cite{Hirsch2001PPCF,Viezzer2013NF}, the $V_{\mathbf{E}_r\times\mathbf{B}}$ is more dominant than $V_{\mathrm{phase}}$ in the edge plasma, which means the $V_{\mathrm{phase}}$ is negligible and then the $E_r$ can be evaluated from $E_r = V_{\perp} B$.


\section{Typical I-mode plasmas under boron coating}

\begin{figure}[htbp]
\centering
\includegraphics[width=0.60\columnwidth]{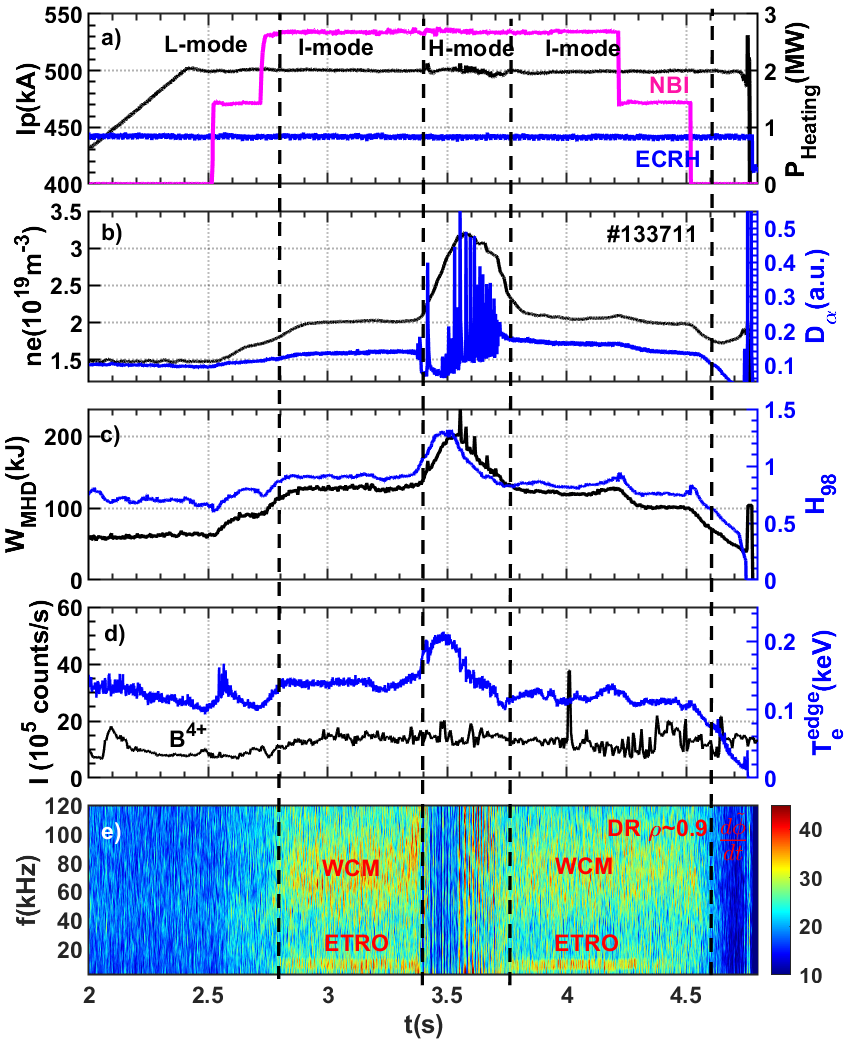}
\caption{The typical experimental results of I-mode plasmas with boron coating. a) the plasma current $I_p$ and the auxiliary heating power; b) the chord-averaged density $\bar{n}_e$ and the line $\mathrm{D}_\alpha$ emission; c) the plasma stored energy $W_{\mathrm{MHD}}$ and $H_{98}$; d) the detector counts in an EUV spectral band corresponding to $\mathrm{B}^{4+}$ radiation and the edge $T_e$; e) the time-frequency spectrogram of the DR phase derivative signal.}
\label{fig:typical}
\end{figure}

The typical experimental results of I-mode regime with boronization (shot\#133711) are displayed in figure~\ref{fig:typical}, with $I_p \sim 500$\,kA, $P_{\mathrm{ECRH}}\sim 0.8$\,MW, $P_{\mathrm{NBI}}\sim 1.4$--$2.7$\,MW. The LSN configuration was preferentially employed for I-mode under boronization experiments during this campaign, because the upper divertor water-cooling system was not operational. The toroidal field was set to $B_t = +2.47$\,T, with $\mathbf{B}\times\nabla B$ points upward, i.e.\ unfavorable-configuration. As the NBI heating power increased, the L--I transition at $t \sim 2.7$\,s was identified by the appearance of the WCM with frequency of 50--100\,kHz and the ETRO with frequency of 9\,kHz in the $\mathrm{d}\tilde{\varphi}/\mathrm{d}t$ spectrogram, as well as the increase of the edge electron temperature, the plasma stored energy $W_{\mathrm{MHD}}$ and the $H_{98}$ factor ($H_{98} = \tau_E / \tau_{\mathrm{IPB98(y,2)}}$)~\cite{ITER1999NF}. Throughout the discharge, the wall condition was boronization, which was consistent with the evolution of the $\mathrm{B}^{4+}$ radiation intensity, as shown in figure~\ref{fig:typical}(d). The I-mode plasma transitioned to the H-mode plasma at $t \sim 3.38$\,s, which was identified by the sudden decrease of the $\mathrm{D}_\alpha$ intensity, as well as the increase of the chord averaged density, the edge temperature, the plasma stored energy, and the $H_{98}$ factor. Subsequently, as the ELM erupted, large amounts of particles and energy were emitted outward, causing the H-mode to collapse, and then the plasma returned to I-mode at $3.78$\,s.

\begin{figure}[htbp]
\centering
\includegraphics[width=0.48\columnwidth]{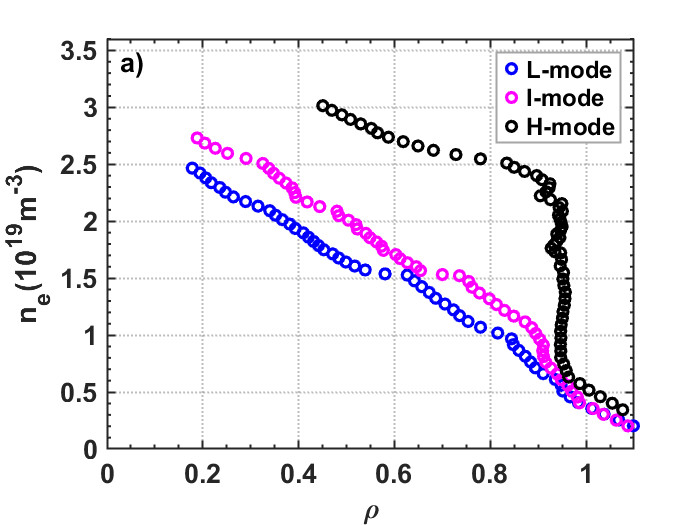}\hfill
\includegraphics[width=0.48\columnwidth]{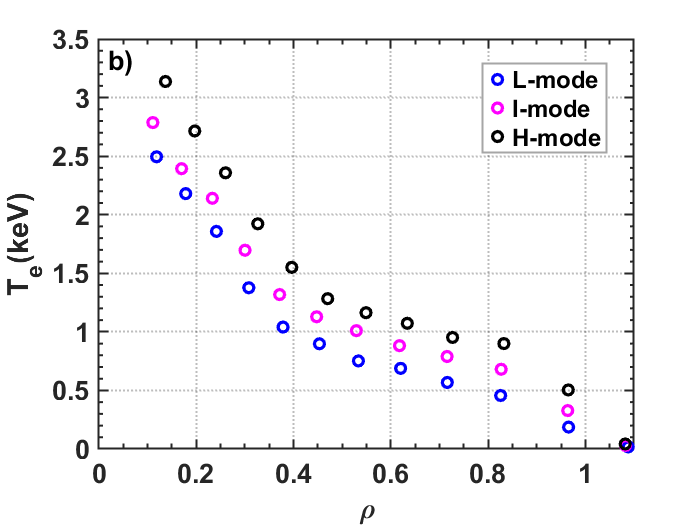}
\caption{Profiles of electron density (a) measured by reflectometry and electron temperature (b) measured by ECE during the L-mode, I-mode, and the H-mode, respectively.}
\label{fig:profiles}
\end{figure}

Figure~\ref{fig:profiles} presents the radial profiles of electron density and electron temperature for L-mode ($4.6$\,s), I-mode ($2.9$\,s), and H-mode ($3.5$\,s). The electron density was measured by a profile reflectometer. In the absence of Thomson scattering data, the electron temperature was obtained from the electron cyclotron emission (ECE) diagnostic. In the figure, raw data are denoted by circles. 
The I-mode density profile resembles that of L-mode and, in contrast to H-mode, does not exhibit a density pedestal. Moreover, compared with L-mode, I-mode shows a clear temperature pedestal, and its edge temperature lies between those of L-mode and H-mode. These features are consistent with the macroscopic I-mode profile characteristics previously observed on EAST.

\begin{figure}[htbp]
\centering
\includegraphics[width=0.48\columnwidth]{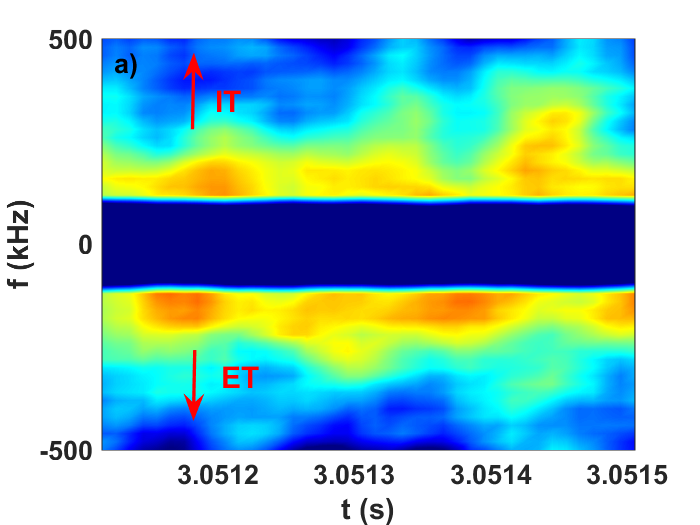}\hfill
\includegraphics[width=0.48\columnwidth]{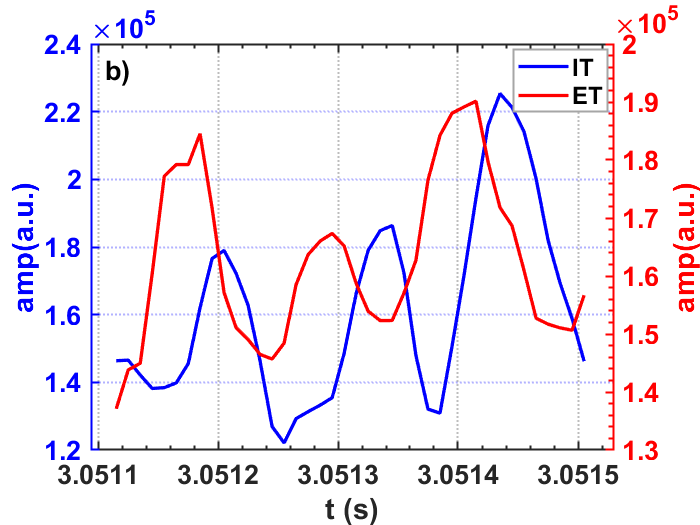}
\caption{Turbulence $A_{ei\varphi}$ spectrogram near the pedestal ($\rho \sim 0.9$) during I-mode (a); (b) the evolution of the electron turbulence (ET) intensity, where the integrating frequency range is ($-0.5$\,MHz, $-0.1$\,MHz) and the ion turbulence (IT) intensity, where the integrating frequency range is ($0.1$\,MHz, $0.5$\,MHz).}
\label{fig:turbulence}
\end{figure}

The turbulence spectrogram measured by DR during the I-mode with boronization (shot \#133711) is shown in figure~\ref{fig:turbulence}. 
In the turbulence spectrogram, ET denotes the turbulence propagating in the electron diamagnetic drift direction, and IT denotes the turbulence propagating in the ion diamagnetic drift direction.
The temporal evolution of the ET and IT amplitudes during the I-mode phase is shown in figure~\ref{fig:turbulence}(b). Consistent with previous results, quasi-periodic transitions between ET and IT are always observed during the presence of ETRO. These turbulence transitions are most likely caused by a competition mechanism between the ion temperature gradient (ITG) mode and the trapped electron mode (TEM). The same microphysical mechanism of ETRO under both boron coating and lithium coating is thus indirectly confirmed, consistent with the ETRO characterization reported in Refs.~\cite{ADLiu2020NF,ADLiu2024NF}.


\section{Statistical comparison of I-mode under boron coating and lithium coating}

\begin{figure}[htbp]
\centering
\includegraphics[width=0.48\columnwidth]{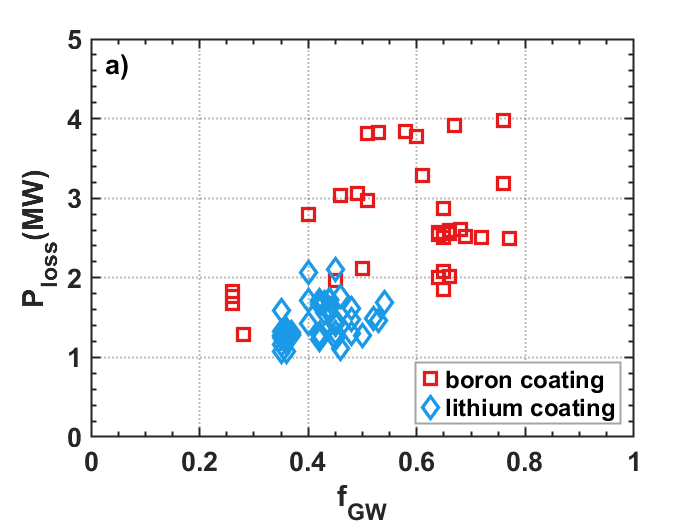}\hfill
\includegraphics[width=0.48\columnwidth]{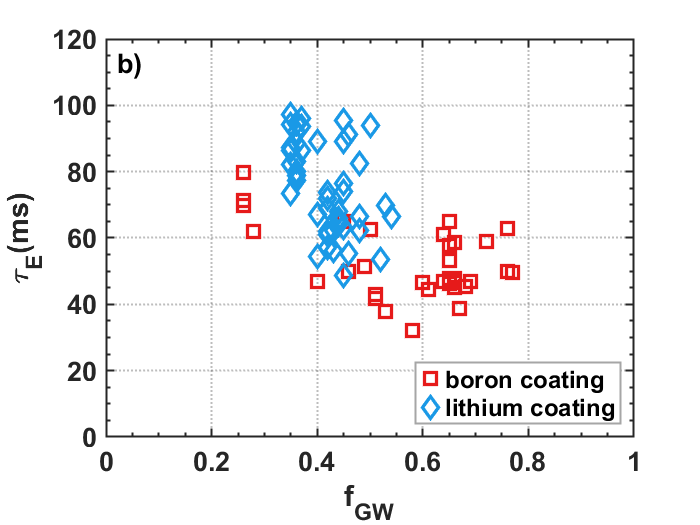}
\caption{Statistical overview of Greenwald density fraction $f_{\mathrm{GW}}$ versus the loss power $P_{\mathrm{loss}}$ a), and the energy confinement time $\tau_E$ b) for I-mode under boron coating and lithium coating.}
\label{fig:stats_density}
\end{figure}

Figure~\ref{fig:stats_density} shows the correlation between the Greenwald density fraction $f_{\mathrm{GW}}$, the loss power $P_{\mathrm{loss}}$, and the energy confinement time $\tau_E$ in I-mode discharges under boron coating and lithium coating. Here, $f_{\mathrm{GW}} = \bar{n}_e / n_{\mathrm{GW}}$, $n_{\mathrm{GW}} = I_p / (\pi a^2)$\cite{Greenwald1988NF}, $P_{\mathrm{loss}} = P_{\mathrm{heat}} - \mathrm{d}W_{\mathrm{MHD}}/\mathrm{d}t$, $\tau_E = W_{\mathrm{MHD}} / P_{\mathrm{loss}}$, where $\bar{n}_e$ is the chord-averaged density (in units of $10^{19}\,\mathrm{m}^{-3}$), $I_p$ is the plasma current (in MA), $a$ is the plasma minor radius (in m), $P_{\mathrm{heat}}$ is the heating power, and $W_{\mathrm{MHD}}$ is the stored energy. The statistical data indicate that for I-mode with lithium coating, the operational density range is limited, with $f_{\mathrm{GW}}$ between 0.3 and 0.5, and $P_{\mathrm{loss}}$ is lower than that of the boron coating case, whereas $\tau_E$ is higher. Notably, the accessible density range is substantially wider under boronization, spanning $f_{\mathrm{GW}}$ from 0.2 to 0.8, which indicates that the boron coating substantially expands the operational density window of I-mode. Furthermore, as $P_{\mathrm{loss}}$ increases, $f_{\mathrm{GW}}$ is observed to gradually increase, while $\tau_E$ exhibits a decreasing trend. This suggests that with increasing density, the power required to access I-mode rises accordingly; meanwhile, since radiation scales with the square of the density, the temperature is reduced, and $\tau_E$ is correspondingly lowered.

\begin{figure}[htbp]
\centering
\includegraphics[width=0.48\columnwidth]{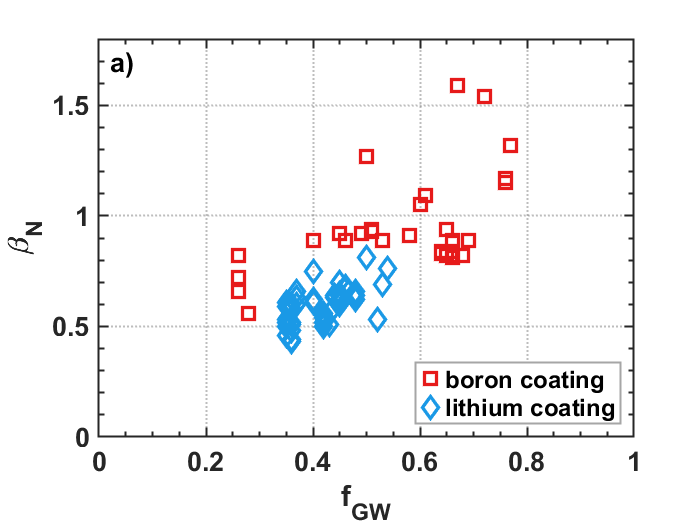}\hfill
\includegraphics[width=0.48\columnwidth]{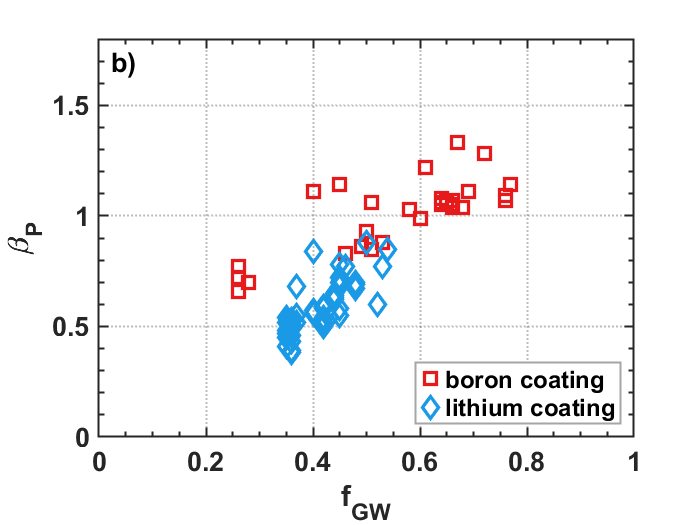}
\caption{Statistical overview of Greenwald density fraction $f_{\mathrm{GW}}$ versus the normalized beta $\beta_N$ a), and poloidal beta $\beta_p$ b) for I-mode under boron coating and lithium coating.}
\label{fig:stats_beta}
\end{figure}

Figure~\ref{fig:stats_beta} presents the statistics of the relationship between the Greenwald density fraction $f_{\mathrm{GW}}$, the normalized beta $\beta_N$, and the poloidal beta $\beta_p$ in I-mode discharges under boron coating and lithium coating. Here, $\beta_N = \beta_t / (I_p / a B_t)$, $\beta_t = 2\mu_0\langle p\rangle / B_t^2$, and $\beta_p = 2\mu_0\langle p\rangle / B_{p}^2$, where $B_t$ is the toroidal magnetic field, $B_p$ is the poloidal magnetic field, $\langle p\rangle$ is the thermal pressure, and $\mu_0$ is the permeability of free space. An approximately linear positive correlation between $f_{\mathrm{GW}}$, $\beta_N$ and $\beta_p$ is observed in both boronization and lithiation I-mode discharges, indicating that the bootstrap current fraction increases with density. This trend of $\beta$ rising with density is similar to that observed for I-mode on AUG~\cite{Ryter2017NF}, suggesting that I-mode may possess a self-consistent relationship that enables the plasma to sustain higher $f_{\mathrm{GW}}$ values at high density. For a reactor, high-density I-mode can achieve higher $\beta_N$ while increasing $f_{\mathrm{GW}}$ --- given that DEMO requires the coexistence of high $\beta_N$ and high $f_{\mathrm{GW}}$ --- which implies that I-mode is a favorable operational regime for DEMO. The observed positive correlation is statistical; the underlying causal physical mechanism --- whether it is the stiffness of the edge transport barrier or the self-organized behavior of core confinement --- remains to be clarified through further investigation.

\begin{figure}[htbp]
\centering
\includegraphics[width=0.48\columnwidth]{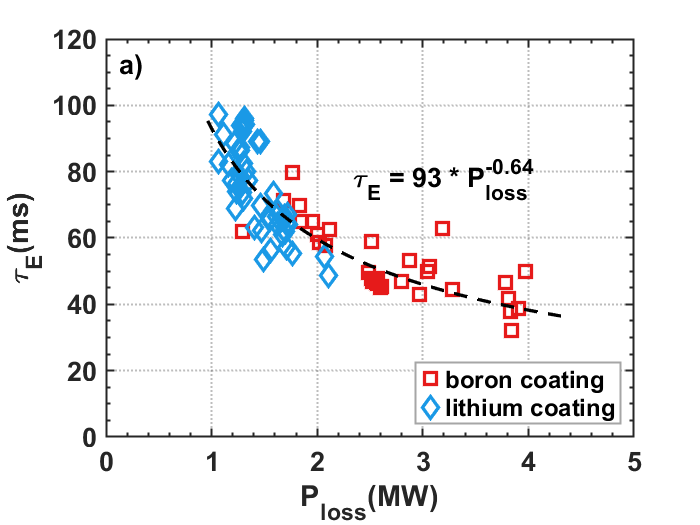}\hfill
\includegraphics[width=0.48\columnwidth]{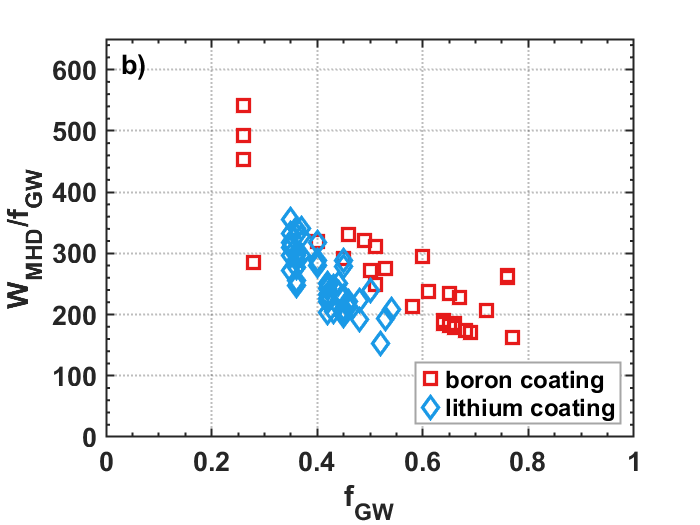}
\caption{Statistical overview of the loss power $P_{\mathrm{loss}}$ versus the energy confinement time $\tau_E$ a) and Greenwald density fraction $f_{\mathrm{GW}}$ versus the ratio $W_{\mathrm{MHD}}/f_{\mathrm{GW}}$ b) for I-mode under boron coating and lithium coating.}
\label{fig:stats_ploss}
\end{figure}

Figure~\ref{fig:stats_ploss} illustrates the relationships between the loss power $P_{\mathrm{loss}}$ and the energy confinement time $\tau_E$, and between the Greenwald density fraction $f_{\mathrm{GW}}$ and the ratio $W_{\mathrm{MHD}}/f_{\mathrm{GW}}$ in I-mode discharges under boron coating and lithium coating. An inverse correlation between $P_{\mathrm{loss}}$ and $\tau_E$ is observed, and the fitted inverse relationship is given by $\tau_E = 93 \times P_{\mathrm{loss}}^{-0.64}$. Furthermore, the ratio $W_{\mathrm{MHD}} / f_{\mathrm{GW}}$ can be regarded as a rough proxy for the average temperature $\bar{T}$. As shown in the figure, $W_{\mathrm{MHD}} / P_{\mathrm{loss}}$ decreases with increasing density, which indicates that although the heating power and the stored energy increase with density, the average temperature decreases. This suggests that confinement is not significantly improved and $\tau_E$ correspondingly decreases with density. The statistical result for $\tau_E$ is mutually consistent with the energy confinement time statistics presented above.

\begin{figure}[htbp]
\centering
\includegraphics[width=0.6\columnwidth]{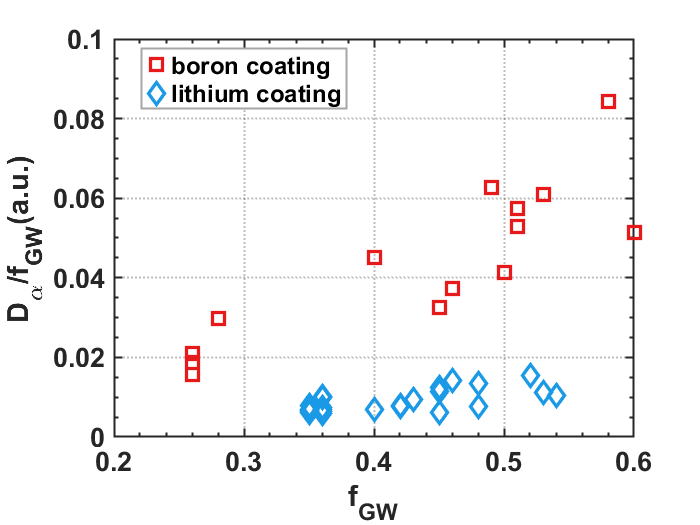}
\caption{Statistical overview of the Greenwald density fraction $f_{\mathrm{GW}}$ versus the ratio of $\mathrm{D}_\alpha /f_{\mathrm{GW}}$ for I-mode under boron coating and lithium coating.}
\label{fig:stats_dalpha}
\end{figure}

As indicated by the preceding statistics, a substantial expansion of the I-mode operational density range is observed with the boron coating compared with the lithium coating. This expansion is likely correlated with the difference in boundary conditions between the two wall treatments. For this purpose, the $\mathrm{D}_\alpha$ emission is compared between boronization and lithiation I-mode discharges under the same toroidal magnetic field and similar density. A higher $\mathrm{D}_\alpha /f_{\mathrm{GW}}$ is observed for the I-mode under boronization than for the I-mode under lithiation, which indirectly suggests a higher level of edge recycling in the boron coating case. This implies a possible physical mechanism for the expanded density range of I-mode with the boron coating: owing to the lower retention of boron compared with lithium, the edge recycling level under boron coating may be higher, thereby raising the edge density and, in turn, effectively increasing the overall chord-averaged density of the plasma. Moreover, many other factors can affect the operational density range, such as the quantitative effect of the boron coating on the L--H transition power threshold, changes in the first-wall/edge temperature, and the fueling/degassing dynamics of the boron coating. All of these will be further validated in depth through dedicated power-scan experiments and edge diagnostics in future work.


\section{I-mode with favorable-configuration under boronization}

I-mode is generally accessed predominantly in the unfavorable-configuration. In the statistics of I-mode under lithiation on EAST, the favorable-configuration is observed in only 8\% of the discharges (4 out of 48 shots), consistent with the rarity or absence of favorable-configuration I-mode reported on Alcator C-Mod~\cite{Hubbard2011POP} and AUG~\cite{Ryter2017NF}. 
In contrast, in the statistics of I-mode under boronization on EAST, a sharp increase in the fraction of favorable-configuration discharges to 51\% (19 out of 37 shots) is found.

\begin{figure}[htbp]
\centering
\includegraphics[width=0.48\columnwidth]{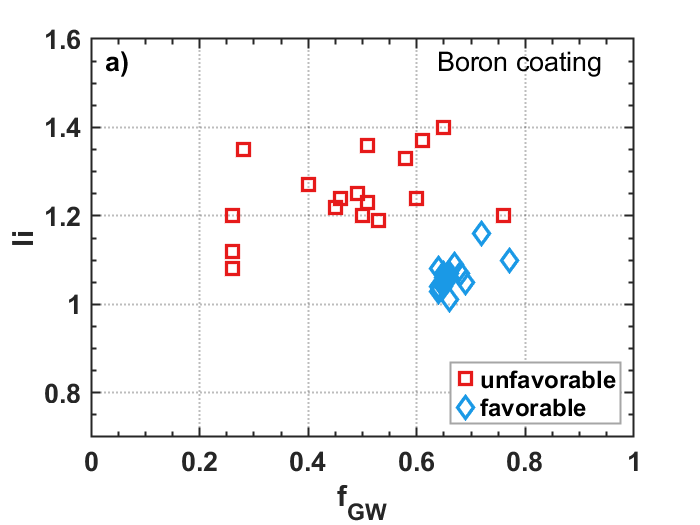}\hfill
\includegraphics[width=0.48\columnwidth]{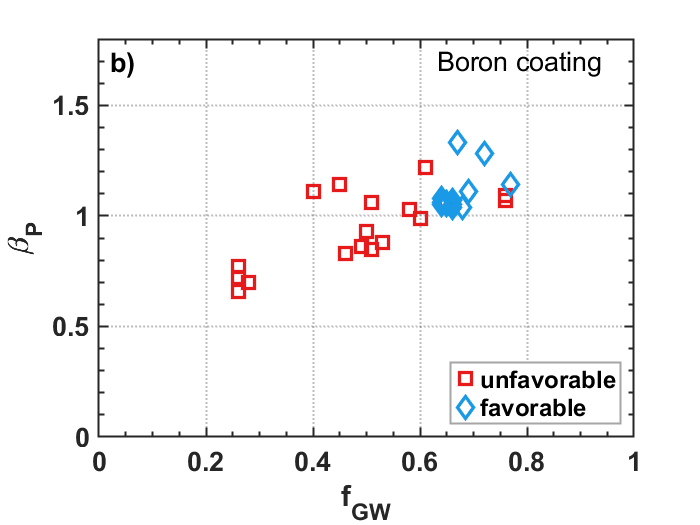}
\caption{Statistical overview of Greenwald density fraction $f_{\mathrm{GW}}$ versus the internal inductance $l_i$ a), and the poloidal beta $\beta_p$ b) for I-mode under boronization with the favorable and the unfavorable configurations.}
\label{fig:fav_stats}
\end{figure}

Figure~\ref{fig:fav_stats} presents the correlation between the Greenwald density fraction $f_{\mathrm{GW}}$ and the internal inductance $l_i$, and the poloidal beta $\beta_p$, in the I-mode under boronization discharges for both favorable and unfavorable configurations. In the favorable-configuration, $f_{\mathrm{GW}}$ is found to lie in the high-density range of $0.6$--$0.8$. This may be attributed to the density dependence of the L--H power threshold: as the density increases, the L--H threshold rises (consistent with the Martin scaling law~\cite{Martin2008JPCS}), thereby broadening the power window between the L--I and I--H transitions and providing operational space for I-mode to exist in the favorable-configuration. At low density, the L--H threshold for the favorable-configuration is lower, and the plasma tends to transition directly from L-mode to H-mode without passing through the I-mode phase. Furthermore, in the favorable-configuration I-mode, the internal inductance $l_i$ is lower and $\beta_p$ is higher compared with the unfavorable-configuration case. These differences are consistent with the high-density feature of the favorable-configuration: the increase in density leads to enhanced plasma thermal pressure and, consequently, a higher $\beta_p$. The reduction in $l_i$ may originate from a broadening of the current profile caused by the increased collisionality at high density, although this remains speculative and requires dedicated current-profile measurements for confirmation. Additionally, it is consistent with the statistical finding that a high $\beta_p$ at high density corresponds to a high bootstrap current fraction. It should be added that, although the non-inductive current drive from LHW/ECRH may also play a role, more in-depth investigation is required in future work.

\begin{figure}[htbp]
\centering
\includegraphics[width=0.48\columnwidth]{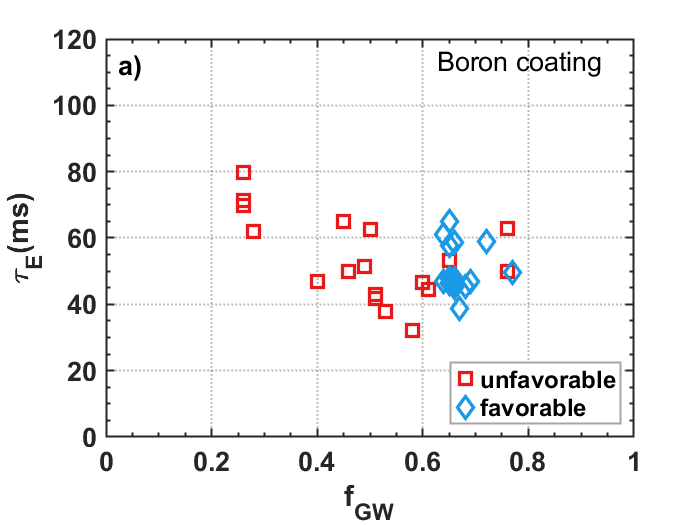}\hfill
\includegraphics[width=0.48\columnwidth]{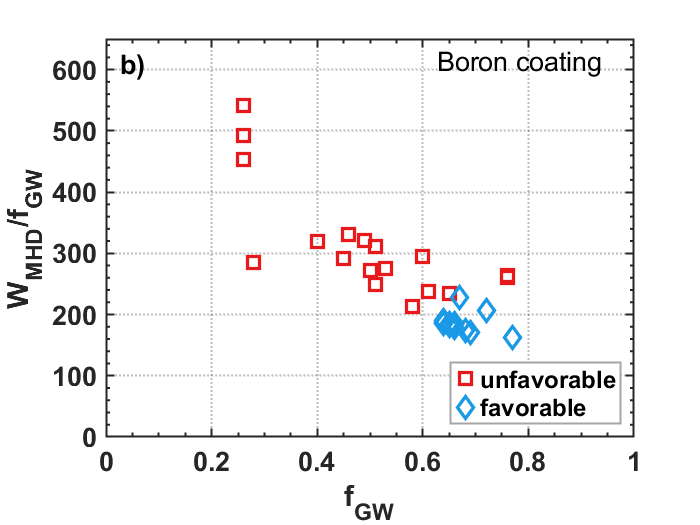}
\caption{Statistical overview of Greenwald density fraction $f_{\mathrm{GW}}$ versus the energy confinement time $\tau_E$ a), and the ratio $W_{\mathrm{MHD}}/f_{\mathrm{GW}}$ b) for I-mode under boronization with the favorable and the unfavorable configurations.}
\label{fig:fav_confinement}
\end{figure}

The relationships between $f_{\mathrm{GW}}$, $\tau_E$, and $W_{\mathrm{MHD}}/f_{\mathrm{GW}}$ for I-mode under boronization in both favorable and unfavorable configurations are presented in figure~\ref{fig:fav_confinement}. Under boron coating, no significant correlation between the energy confinement time and the configuration is observed. However, the ratio $W_{\mathrm{MHD}}/f_{\mathrm{GW}}$, which is taken as a rough proxy for the average temperature $\bar{T}$, is found to be lower in the favorable-configuration than in the unfavorable one, which may be attributed to the higher density in the favorable-configuration.

\begin{figure}[htbp]
\centering
\includegraphics[width=0.6\columnwidth]{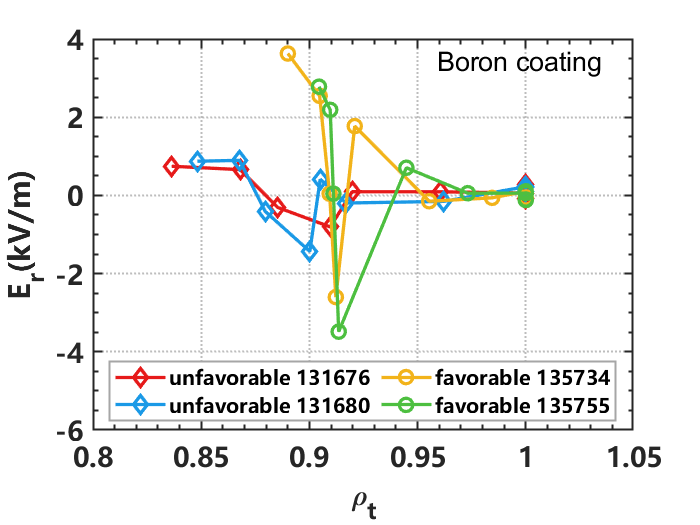}
\caption{Radial electric field $E_r$ profiles of I-mode under boronization in favorable and unfavorable configurations.}
\label{fig:Er_profiles}
\end{figure}

A comparison of the radial electric field $E_r$ in I-mode under boronization between favorable and unfavorable configurations is presented in figure~\ref{fig:Er_profiles}. For this comparison, four discharges with the same plasma current of $I_p = 0.4$\,MA and similar chord-averaged densities close to $\bar{n}_e \sim 4.0\times 10^{19}\,\mathrm{m}^{-3}$ are selected from the two configurations. On the assumption that the turbulent phase velocity is much smaller than the $\mathbf{E_r}\times\mathbf{B}$ flow velocity, $E_r$ is estimated from the poloidal velocity of the plasma measured by DR in the laboratory frame. As shown in figure~\ref{fig:Er_profiles}, a deeper $E_r$ well and a larger maximum $E_r$ shear are observed in the favorable-configuration than in the unfavorable one. In addition, the radial position of the deepest point of the $E_r$ well is found to be located further outward in the favorable-configuration. The above observations may be explained by the high density in the favorable-configuration, which raises the edge pressure gradient and thereby enhances $E_r$. As a result, the stronger $E_r$ shear is able to suppress the enhanced edge turbulence driven by the larger free energy at high density, thereby enabling the L--I transition.

\section{EAST I-mode energy confinement time scaling law}

\begin{table}[htbp]
	\caption{Statistical summary of the two I-mode datasets used for the confinement scaling analysis.}
	\label{tab:dataset_summary}
	\begin{indented}
		\item[]\begin{tabular}{lcc}
			\br
			Quantity & Lithium-conditioned I-mode & Boronized-wall I-mode \\
			\mr
			Number of shots & 48 & 37 \\
			$B_t$ (T) & 2.47 & 2.47 \\
			$I_p$ range (MA) & 0.45--0.60 & 0.40--0.50 \\
			$P_{\mathrm{loss}}$ range (MW) & 1.0--2.0 & 1.3--3.0 \\
			$\bar{n}_e$ range ($10^{19}\,\mathrm{m}^{-3}$) & 2.8--4.2 & 1.8--6.0 \\
			$f_{\mathrm{GW}}$ range & 0.35--0.54 & 0.26--0.77 \\
			Favorable/unfavorable cases & 4/44 & 19/18 \\
			$\tau_E$ range (ms) & 49--97 & 32--80 \\
			$H_{98}$ range & 0.60--1.10 & 0.68--1.00 \\
			\br
		\end{tabular}
	\end{indented}
\end{table}

\begin{figure}[htbp]
	\centering
	\includegraphics[width=0.65\columnwidth]{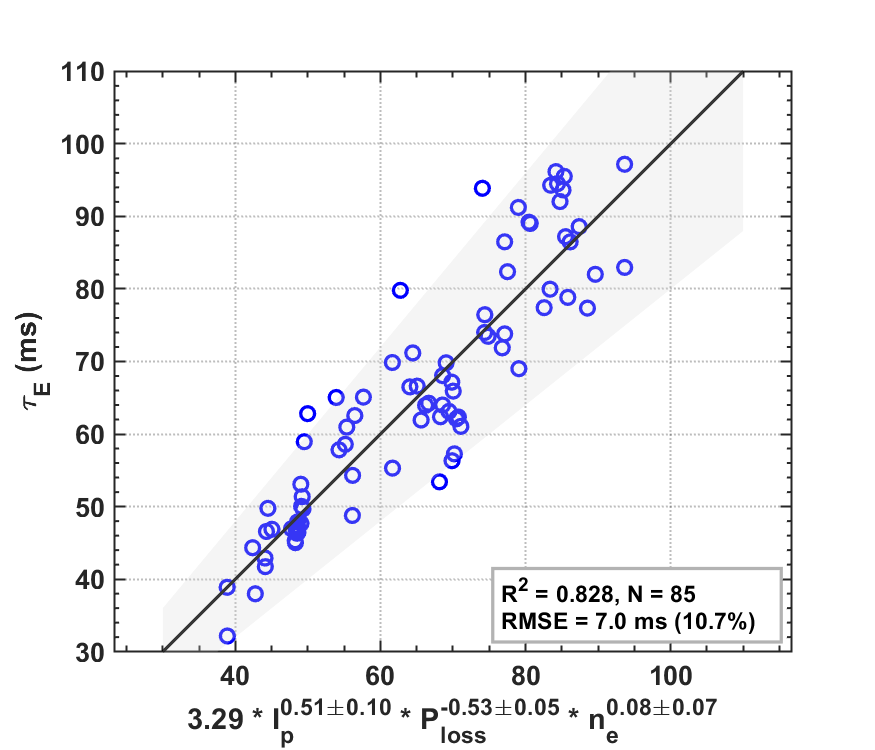}
	\caption{EAST I-mode experimental energy confinement time versus power law scaling calculated from a multiple regression fit for parameters $I_p$, $P_{\mathrm{loss}}$, and $\bar{n}_e$.}
	\label{fig:scaling}
\end{figure}

Having characterized the I-mode operational space under both wall conditions and configuration, we now derive an empirical energy confinement scaling. 
Table~\ref{tab:dataset_summary} provides a statistical overview of the two datasets used in the regression. 
Because all I-mode under boronization discharges on EAST were obtained at a single toroidal field of $B_t = 2.47$\,T, the lithium coating data are likewise restricted to this field; $B_t$ is therefore not included as a regression variable. The scaling is established by multiple log-linear regression (log-OLS) on 85 stationary time windows, of which 37 are taken from I-mode shots under boronization and 48 from I-mode shots under lithiation. 
The energy confinement time $\tau_E$ is taken as the dependent variable, and the plasma current $I_p$, the loss power $P_{\mathrm{loss}}$, and the chord-averaged electron density $\bar{n}_e$ are taken as the independent variables. Therefore, a power-law form is assumed for $\tau_E$, i.e., $\tau_E = c_0 I_p^{c_1} P_{\mathrm{loss}}^{c_2} \bar{n}_e^{c_3}$. Natural logarithms are taken on both sides, transforming the equation into a multiple linear regression problem: $\ln(\tau_E) = \ln(c_0) + c_1 \ln(I_p) + c_2 \ln(P_{\mathrm{loss}}) + c_3 \ln(\bar{n}_e)$. The coefficients are solved by the least-squares method, yielding $c_0 = 3.29$, $c_1 = 0.51$, $c_2 = -0.53$, $c_3 = 0.08$, which gives
\begin{equation}
	\tau_{E, EAST, I-mode} = 3.29 \, I_p^{0.51 \pm 0.10} \, P_{\mathrm{loss}}^{-0.53 \pm 0.05} \, \bar{n}_e^{0.08 \pm 0.07} \qquad (B_t = 2.47\,\mathrm{T}).
	\label{eq:scaling}
\end{equation}
The exponents of $I_p$ and $P_{\mathrm{loss}}$ are found to be highly significant, whereas the exponent of $\bar{n}_e$ is not significant, indicating that the energy confinement in I-mode is mainly determined by the plasma current and the loss power, with a very weak density dependence. For the model, $R^2 = 0.828$ and $\mathrm{RMSE} = 7.0$\,ms are obtained in linear space (the mean $\tau_E$ being about $65$\,ms, corresponding to a relative error of approximately 11\%). Compared with the IPB98(y,2) H-mode scaling~\cite{ITER1999NF}, the I-mode confinement exhibits a much weaker dependence on $P_{\mathrm{loss}}$ and a slower power degradation, which is more favorable for future fusion reactors. Furthermore, the confinement is almost unaffected by density, a feature similar to that found in the C-Mod I-mode scaling~\cite{Wilks2019}. It should be noted that in the bivariate fit $\tau_E = 93 \times P_{\mathrm{loss}}^{-0.64}$, the power exponent ($-0.64$) is more negative than the exponent in the multivariate fit ($-0.53$). This difference arises because a positive correlation exists between $I_p$ and $P_{\mathrm{loss}}$ --- that is, higher-power discharges usually correspond to higher plasma currents. When the $I_p$ term is included in the multivariate regression, the independent contribution of $P_{\mathrm{loss}}$ is better separated; therefore, the multivariate scaling $\tau_{E, EAST, I-mode} = 3.29 I_p^{0.51 \pm 0.10} P_{\mathrm{loss}}^{-0.53 \pm 0.05} \bar{n}_e^{0.08 \pm 0.07}$ is considered more reliable. In addition, compared with the C-Mod I-mode scaling ($\tau_E \propto I_p^{0.69} P_{\mathrm{loss}}^{-0.29}$), the EAST I-mode confinement shows a weaker dependence on $I_p$ (0.51 vs.\ 0.69) and a steeper power degradation ($-0.53$ vs.\ $-0.29$). A feature common to both devices is that the confinement is almost independent of density (exponents of 0.08 and 0.02, respectively). This may be attributed to the absence of a density pedestal in I-mode: the core density is mainly determined by edge particle transport rather than by a pedestal, so the contribution of density to confinement is limited. Possible reasons for the steeper power degradation exponent in EAST I-mode compared with C-Mod are: (1) EAST I-mode is dominantly heated by LHW/ECRH, and the wave-plasma coupling efficiency is correlated with the heating power; (2) the plasma current ($0.4$--$0.5$\,MA) and toroidal magnetic field ($2.47$\,T) of EAST I-mode are both significantly lower than those of C-Mod ($I_p$ $0.55$--$1.7$\,MA, $B_t$ $2.8$--$8$\,T), and differences in the dimensionless parameter space may affect the confinement scaling. 
Nevertheless, the EAST I-mode power degradation ($-0.53$) remains weaker than the H-mode IPB98(y,2) value of $-0.69$, confirming that I-mode confinement degrades more slowly with power.


\section{Discussion}

In this section, we discuss three aspects of the I-mode under boronization that merit further consideration.

\subsection{Expanded density window with boronization}

A substantially wider operational density range is observed for the I-mode under boronization ($f_{\mathrm{GW}}=0.26$--$0.77$) than for the I-mode under lithiation ($f_{\mathrm{GW}}=0.35$--$0.54$). A possible cause for this difference is the distinct edge recycling levels under the two wall conditions. Owing to the lower retention of hydrogen isotopes in boron compared with lithium, a higher level of edge neutral recycling is produced under the boron coating, which raises the edge density. The statistical results of $\mathrm{D}_\alpha$ in figure~\ref{fig:stats_dalpha} indirectly support this explanation: a higher $\mathrm{D}_\alpha$ is observed for the I-mode under boronization, indicating that the edge recycled particles contribute to more efficient fueling rather than mere radiation loss. In addition, the increased recycling level is also responsible for a higher L--H transition power threshold~\cite{Battaglia2013NF,Zuo2012PPCF}, thereby broadening the power window for I-mode and enabling its sustainment at higher densities without an I--H transition. Furthermore, the quantitative impact of the boron coating on the L--I and I--H transition thresholds still needs to be clarified through dedicated power scan experiments in future work.

\subsection{Formation mechanism of favorable-configuration I-mode}

It is generally accepted that a lower L--H power threshold is typically associated with the favorable-configuration, which makes a direct L--H transition more likely and consequently leaves a narrow power window for I-mode. A striking contrast is observed between the high fraction ($\sim$51\%) of favorable-configuration discharges in the I-mode under boronization and the extreme rarity (only 4 shots, $\sim$8\%) of such discharges under lithium coating. The appearance of favorable-configuration I-mode under the boron coating is inferred to be closely linked to the density increase: as the density rises, the L--H transition power threshold increases, thereby broadening the power window between the L--I and I--H transitions and providing the necessary condition for I-mode to exist in the favorable-configuration. This also explains why all the observed boron coating favorable I-modes are concentrated in the higher density range ($f_{\mathrm{GW}}=0.6$--$0.77$). At low density, the L--H threshold for the favorable-configuration is lower, and the plasma is more prone to a direct transition into H-mode. Notably, a deeper radial electric field well and a stronger $\mathbf{E_r}\times\mathbf{B}$ velocity shear are exhibited by the favorable-configuration I-mode (figure~\ref{fig:Er_profiles}), implying that an enhanced turbulence suppression capability is required to balance the increased free energy associated with the higher density. This phenomenon is qualitatively consistent with the physical picture derived by Cziegler from nonlinear three‑wave coupling analysis on C‑Mod, where a stronger energy transfer from turbulence to zonal flows is found in the favorable-configuration~\cite{Cziegler2017PRL}.

\subsection{Steady-state I-mode sustainment and ETRO occurrence frequency}

Under boron coating, three of the four I-mode types classified by~\cite{Zhong2024NF} are observed. These are Type I (GAM + WCM), Type II (ETRO + WCM) and Type IV (WCM only), comprising 5, 6 and 26 of 37 discharges, respectively. Notably, Type III is absent from the boron coating dataset. 
Previous work has demonstrated that ETRO sustains steady-state I-mode by modulating quasi-periodic transitions between electron and ion turbulence in I-mode under lithiation, where ETRO is indeed observed in most steady-state discharges on EAST~\cite{ADLiu2020NF,ADLiu2024NF}. 
Under boron coating, steady-state I-mode can be maintained by multiple turbulence regimes. Here, we define steady-state I-mode operation as discharges whose I-mode duration is limited only by the plasma flat-top length. Both Type I (GAM + WCM) and Type IV (WCM only) can sustain stationary I-mode under this definition, indicating that either GAM or WCM alone may provide sufficient edge turbulence regulation.  
In addition, it is observed that ETRO occurs less frequently in I-mode under boronization than in I-mode under lithiation.
The reduced occurrence of ETRO under boron coating (only $\sim$15\% of discharges) may be attributed to the lower edge electron temperature $T_e$ in these plasmas, which is suggested by the lower average energy confinement time (51.9\,ms for boron coating vs.\ 75.2\,ms for lithium coating). The presence or absence of ETRO may therefore reflect systematic differences in the edge temperature gradient between the two wall conditions. 
This hypothesis warrants testing through higher-spatiotemporal-resolution $T_e$ diagnostics and gyrokinetic simulations in future work. 
Boronized-wall I-mode on EAST exhibits a wider density range and unexpectedly frequent favorable-configuration access, suggesting a promising route toward reactor-relevant ELM-free operation, while dedicated threshold and recycling studies are required to establish the underlying causality.


\section{Summary and outlook}

We have presented the first systematic study of I-mode operation under boronized wall conditions on EAST and compared it with a lithium-conditioned I-mode database obtained at the same toroidal field, $B_t = 2.47$\,T. 
The results show that boronized-wall I-mode substantially broadens the accessible density range and enables frequent access to favorable-configuration I-mode. 

First, the accessible Greenwald fraction is extended from $f_{\mathrm{GW}} = 0.35 - 0.54$ under lithiation to $f_{\mathrm{GW}} = 0.26 - 0.77$ under boronization. The higher normalized $\mathrm{D}_\alpha$ emission observed in the boronized-wall dataset suggests that enhanced edge recycling may contribute to the increased density access. Dedicated matched power, density and fueling scans will be required to quantify the relative roles of wall recycling, fueling efficiency and transition thresholds. 

Second, favorable-configuration I-mode is observed much more frequently under boronization, accounting for 19 of 37 discharges ($51\%$), compared with only 4 of 48 discharges under lithiation ($8\%$). These favorable-configuration cases occur predominantly at high density, $f_{\mathrm{GW}} = 0.6 - 0.77$, and are characterized by lower internal inductance, higher poloidal beta, a deeper radial electric-field well and stronger $\mathbf{E_r}\times\mathbf{B}$ velocity shear. These observations suggest that the high-density condition may raise the L–H transition threshold and thereby broaden the power window in which favorable-configuration I-mode can be sustained.

Third, when ETRO is present under boronization, the associated quasi-periodic transition between electron and ion turbulence is similar to that observed under lithiation. However, ETRO is detected in only a minority of boronized-wall I-mode discharges. Steady I-mode operation can also be sustained in Type I (GAM + WCM) and Type IV (WCM only) regimes, indicating that ETRO is not the only route to stationary I-mode on EAST. 

Finally, an empirical EAST I-mode confinement scaling at fixed toroidal field has been obtained:  $\tau_{E, EAST, I-mode} = 3.29 I_p^{0.51 \pm 0.10} P_{\mathrm{loss}}^{-0.53 \pm 0.05} \bar{n}_e^{0.08 \pm 0.07}$.
The scaling shows a weaker power degradation than the IPB98(y,2) H-mode scaling and an almost negligible density dependence, consistent with the absence of a density pedestal in I-mode. 

Future experiments should focus on matched lithium/boron/coating-free comparisons, dedicated L–I and I–H threshold scans, direct measurements of recycling and neutral fueling, and high-resolution edge profile measurements. Such studies will be essential for assessing whether high-density, favorable-configuration I-mode can be robustly extrapolated to full-metal-wall, reactor-relevant conditions.


\section*{Acknowledgments}

This work was supported by the National MCF Energy R\&D Program 2024YFE03200001, the National Natural Science Foundation of China under Contract No.12375230, and the Youth Innovation Promotion Association CAS No.2023470.

\section*{References}


\bibliographystyle{iopart-num}
\bibliography{Imode_boronization}

@article{Greenwald1988NF,
  author = {Greenwald, M. and Terry, J.L. and Wolfe, S.M. and Ejima, S. and Bell, M.G. and Kaye, S.M. and Neilson, G.H.},
  title = {A new look at density limits in tokamaks},
  journal = {Nucl. Fusion},
  volume = {28},
  number = {12},
  pages = {2199--2207},
  year = {1988},
  doi = {10.1088/0029-5515/28/12/009}
}

@article{Greenwald1997NF,
  author = {Greenwald, M. and Boivin, R.L. and Bombarda, F. and Bonoli, P.T. and Fiore, C.L. and Garnier, D. and Goetz, J.A. and Golovato, S.N. and Graf, M.A. and Granetz, R.S. and Horne, S. and Hubbard, A. and Hutchinson, I.H. and Irby, J.H. and LaBombard, B. and Lipschultz, B. and Marmar, E.S. and May, M.J. and McCracken, G.M. and O'Shea, P. and Rice, J.E. and Schachter, J. and Snipes, J.A. and Stek, P.C. and Takase, Y. and Terry, J.L. and Wang, Y. and Watterson, R. and Welch, B. and Wolfe, S.M.},
  title = {H mode confinement in {Alcator C-Mod}},
  journal = {Nucl. Fusion},
  volume = {37},
  pages = {793},
  year = {1997},
  doi = {10.1088/0029-5515/37/6/I07}
}

@article{Ryter1998PPCF,
  author = {Ryter, F. and Suttrop, W. and Br\"{u}sehaber, B. and Kaufmann, M. and Mertens, V. and Murmann, H. and Peeters, A.G. and Stober, J. and Zohm, H. and the ASDEX Upgrade Team},
  title = {H-mode power threshold and confinement in ASDEX Upgrade},
  journal = {Plasma Phys. Control. Fusion},
  volume = {40},
  pages = {725},
  year = {1998},
  doi = {10.1088/0741-3335/40/5/028}
}

@article{Whyte2010NF,
  author = {Whyte, D.G. and Hubbard, A.E. and Hughes, J.W. and Lipschultz, B. and Rice, J.E. and Marmar, E.S. and Greenwald, M. and Cziegler, I. and Dominguez, A. and Golfinopoulos, T. and Howard, N.T. and Lin, L. and McDermott, R.M. and Porkolab, M. and Reinke, M.L. and Terry, J. and Tsujii, N. and Wolfe, S. and Zweben, S. and the Alcator C-Mod Team},
  title = {I-mode: an {H}-mode energy confinement regime with {L}-mode particle transport in {Alcator C-Mod}},
  journal = {Nucl. Fusion},
  volume = {50},
  pages = {105005},
  year = {2010},
  doi = {10.1088/0029-5515/50/10/105005}
}

@article{Hubbard2011POP,
  author = {Hubbard, A.E. and Whyte, D.G. and Churchill, R.M. and Cziegler, I. and Dominguez, A. and Golfinopoulos, T. and Hughes, J.W. and Rice, J.E. and Bespamyatnov, I. and Greenwald, M.J. and Howard, N. and Lipschultz, B. and Marmar, E.S. and Reinke, M.L. and Rowan, W.L. and Terry, J.L. and the Alcator C-Mod Group},
  title = {Edge energy transport barrier and turbulence in the {I}-mode regime on {Alcator C-Mod}},
  journal = {Phys. Plasmas},
  volume = {18},
  pages = {056115},
  year = {2011},
  doi = {10.1063/1.3582135}
}

@article{Cziegler2017PRL,
  author = {Cziegler, I. and Hubbard, A.E. and Hughes, J.W. and Terry, J.L. and Tynan, G.R.},
  title = {Turbulence nonlinearities shed light on geometric asymmetry in tokamak confinement transitions},
  journal = {Phys. Rev. Lett.},
  volume = {118},
  pages = {105003},
  year = {2017},
  doi = {10.1103/PhysRevLett.118.105003}
}

@article{Walk2014POP,
  author = {Walk, J.R. and Hughes, J.W. and Hubbard, A.E. and Terry, J.L. and Whyte, D.G. and White, A.E. and Baek, S.G. and Reinke, M.L. and Theiler, C. and Churchill, R.M. and Rice, J.E. and Snyder, P.B. and Osborne, T. and Dominguez, A. and Cziegler, I.},
  title = {Edge-localized mode avoidance and pedestal structure in {I}-mode plasmas},
  journal = {Phys. Plasmas},
  volume = {21},
  pages = {056103},
  year = {2014},
  doi = {10.1063/1.4872220}
}

@article{Marinoni2015NF,
  author = {Marinoni, A. and Rost, J.C. and Porkolab, M. and Hubbard, A.E. and Osborne, T.H. and White, A.E. and Whyte, D.G. and Rhodes, T.L. and Davis, E.M. and Ernst, D.R. and Burrell, K.H. and the DIII-D Team},
  title = {Characterization of density fluctuations during the search for an {I}-mode regime on the {DIII-D} tokamak},
  journal = {Nucl. Fusion},
  volume = {55},
  pages = {093019},
  year = {2015},
  doi = {10.1088/0029-5515/55/9/093019}
}

@article{Hubbard2017NF,
  author = {Hubbard, A.E. and Osborne, T. and Ryter, F. and Austin, M. and Barr, J.L. and Churchill, R.M. and Cziegler, I. and Fenzi, C. and Hughes, J.W. and Lunsford, R. and Marinoni, A. and McDermott, R.M. and Nespoli, F. and Odstr\v{c}il, T. and Paz-Soldan, C. and Reinke, M.L. and Sciortino, F. and Silvagni, D. and Terry, J.L. and Walk, J.R. and Wilks, T. and Wolfe, S. and the Alcator C-Mod Team and the ASDEX Upgrade Team and the DIII-D Team},
  title = {Physics and performance of the {I}-mode regime over an expanded operating space on {Alcator C-Mod}},
  journal = {Nucl. Fusion},
  volume = {57},
  pages = {126039},
  year = {2017},
  doi = {10.1088/1741-4326/aa8570}
}

@article{Ryter2017NF,
  author = {Ryter, F. and Fischer, R. and Fuchs, J.C. and Happel, T. and McDermott, R.M. and Rathgeber, S.K. and Schuhmann, H. and Viezzer, E. and Wolfrum, E. and the ASDEX Upgrade Team},
  title = {{I}-mode at {ASDEX Upgrade}: {L}-{I} and {I}-{H} transitions, pedestal and confinement properties},
  journal = {Nucl. Fusion},
  volume = {57},
  pages = {016004},
  year = {2017},
  doi = {10.1088/0029-5515/57/1/016004}
}

@article{Wilks2019,
  author = {Wilks, T.M. and Gao, X. and Greenwald, M. and Hughes, J.W. and Hubbard, A.E. and Porkolab, M. and Whyte, D.G. and the Alcator C-Mod Team},
  title = {Scaling of energy confinement time in {I}-mode plasmas on {Alcator C-Mod} using dimensionless parameters},
  journal = {Nucl. Fusion},
  volume = {59},
  pages = {126023},
  year = {2019},
  doi = {10.1088/1741-4326/ab40e5}
}

@article{Silvagni2023,
  author = {Silvagni, D. and Pablant, N. and Ploeckl, B. and Bock, A. and Cathey, A. and Dunne, M. and Eich, T. and F\"{u}l\"{o}p, L. and Happel, T. and Hubbard, A.E. and Kallenbach, A. and Lang, P.T. and Luda, T. and Maraschek, M. and McDermott, R.M. and Ryter, F. and Stober, J. and Tardini, G. and Treutterer, W. and Viezzer, E. and the ASDEX Upgrade Team},
  title = {Pellet-fueled {I}-mode plasmas in {ASDEX Upgrade}},
  journal = {Nucl. Fusion},
  volume = {63},
  pages = {084001},
  year = {2023},
  doi = {10.1088/1741-4326/acde8c}
}

@article{Feng2019NF,
  author = {Feng, X. and Liu, A.D. and Zhou, C. and Liu, Z.X. and Wang, M.Y. and Zhuang, G. and Zou, X.L. and Wang, T.B. and Zhang, Y.Z. and Xie, J.L. and Liu, H.Q. and Zhang, T. and Liu, Y. and Duan, Y.M. and Hu, L.Q. and Hu, G.H. and Kong, D.F. and Wang, S.X. and Zhao, H.L. and Li, Y.Y. and Shao, L.M. and Xia, T.Y. and Ding, W.X. and Lan, T. and Li, H. and Mao, W.Z. and Liu, W.D. and Gao, X. and Li, J.G. and Zhang, S.B. and Zhang, X.H. and Liu, Z.Y. and Qu, C.M. and Zhang, S. and Zhang, J. and Ji, J.X. and Fan, H.R. and Zhong, X.M.},
  title = {{I}-mode investigation on the {Experimental Advanced Superconducting Tokamak}},
  journal = {Nucl. Fusion},
  volume = {59},
  pages = {096025},
  year = {2019},
  doi = {10.1088/1741-4326/ab28a7}
}

@article{YJLiu2020NF,
  author = {Liu, Y.J. and Liu, Z.X. and Liu, A.D. and Zhou, C. and Feng, X. and Yang, Y. and Zhang, T. and Xia, T.Y. and Liu, H.Q. and Wu, M.Q. and Zou, X.L. and Kong, D.F. and Li, H. and Xie, J.L. and Lan, T. and Mao, W.Z. and Zhang, S.B. and Ding, W.X. and Zhuang, G. and Liu, W.D. and the EAST Team},
  title = {Power threshold and confinement of the {I}-mode in the {EAST} tokamak},
  journal = {Nucl. Fusion},
  volume = {60},
  pages = {082003},
  year = {2020},
  doi = {10.1088/1741-4326/ab88e0}
}

@article{ADLiu2020NF,
  author = {Liu, A.D. and Zou, X.L. and Han, M.K. and Wang, T.B. and Zhou, C. and Wang, M.Y. and Duan, Y.M. and Verdoolaege, G. and Dong, J.Q. and Wang, Z.X. and Xi, F. and Xie, J.L. and Zhuang, G. and Ding, W.X. and Zhang, S.B. and Liu, Y. and Liu, H.Q. and Wang, L. and Li, Y.Y. and Wang, Y.M. and Lv, B. and Hu, G.H. and Zhang, Q. and Wang, S.X. and Zhao, H.L. and Qu, C.M. and Liu, Z.X. and Liu, Z.Y. and Zhang, J. and Lan, T. and Li, H. and Mao, W.Z. and Liu, W.D.},
  title = {Experimental identification of edge temperature ring oscillation and alternating turbulence transitions near the pedestal top for sustaining stationary {I}-mode},
  journal = {Nucl. Fusion},
  volume = {60},
  pages = {126016},
  year = {2020},
  doi = {10.1088/1741-4326/abb14a}
}

@article{ADLiu2024NF,
  author = {Liu, A.D. and Zou, X.L. and Zhong, X.M. and Song, Y.T. and Han, M.K. and Duan, Y.M. and Liu, H.Q. and Wang, T.B. and Li, E.Z. and Zhang, L. and Feng, X. and Zhuang, G. and the EAST I-mode Working Group},
  title = {Characteristics of edge temperature ring oscillation during the stationary improved confinement mode in {EAST}},
  journal = {Nucl. Fusion},
  volume = {64},
  pages = {016013},
  year = {2024},
  doi = {10.1088/1741-4326/ad0acd}
}

@article{Zhong2022NF,
  author = {Zhong, X.M. and Zou, X.L. and Liu, A.D. and Song, Y.T. and Zhuang, G. and Li, E.Z. and Zhang, B. and Zhang, J. and Zhou, C. and Feng, X. and Duan, Y.M. and Ding, R. and Liu, H.Q. and Lv, B. and Wang, L. and Xu, L.Q. and Zhang, L. and Zhao, H.L. and Zang, Q. and Zhang, T. and Ding, B.J. and Li, M.H. and Qin, C.M. and Wang, X.J. and Zhang, X.J. and the EAST Team},
  title = {Characterization of pedestal burst instabilities during {I}-mode to {H}-mode transition in the {EAST} tokamak},
  journal = {Nucl. Fusion},
  volume = {62},
  pages = {066046},
  year = {2022},
  doi = {10.1088/1741-4326/ac60e9}
}

@article{Zhong2024NF,
  author = {Zhong, X.M. and Zou, X.L. and Liu, A.D. and Song, Y.T. and Zhuang, G. and Liu, H.Q. and Xu, L.Q. and Li, E.Z. and Zhang, B. and Zuo, G.Z. and Wang, Z. and Zhou, C. and Zhang, J. and Shi, W.X. and Gao, L.T. and Wang, S.F. and Gao, W. and Jia, T.Q. and Zang, Q. and Zhao, H.L. and Wang, M. and Xu, H.D. and Wang, X.J. and Gao, X. and Lin, X.D. and Li, J.G. and the EAST I-mode Working Group and the EAST Team},
  title = {{I}-mode plasma confinement improvement by real-time lithium injection and its classification on {EAST} tokamak},
  journal = {Nucl. Fusion},
  volume = {64},
  pages = {126040},
  year = {2024},
  doi = {10.1088/1741-4326/ad80a8}
}

@article{Wang2023PST,
  author = {Wang, P. and Hu, G.H. and Wang, L. and Yan, N. and Zhong, X.M. and Xu, G.S. and Feng, X. and Ye, Y. and Ding, G.F. and Yu, L. and Liu, A.D. and Duan, Y.M. and Li, E.Z. and Xu, L.Q. and Liu, H.Q. and Ding, R. and Zhang, B. and Li, M.H. and Ding, B.J. and Qin, C.M. and Zhang, X.J. and Wang, X.J. and Lyu, B. and Zhang, L. and Wu, M.F. and Zang, Q. and Lin, X. and Zou, X.L. and Song, Y.T.},
  title = {Blob properties in {I}-mode and {ELM}-free {H}-mode compared to {L}-mode on {EAST}},
  journal = {Plasma Sci. Technol.},
  volume = {25},
  pages = {045106},
  year = {2023},
  doi = {10.1088/2058-6272/aca741}
}

@article{Liang2023NF,
  author = {Liang, A.S. and Zou, X.L. and Zhong, W.L. and Xiao, G.L. and Ke, R. and He, X.X. and the HL-2A Team},
  title = {Identification of {I}-mode with ion {ITB} in {NBI}-heated plasmas on the {HL-2A} tokamak},
  journal = {Nucl. Fusion},
  volume = {63},
  pages = {056017},
  year = {2023},
  doi = {10.1088/1741-4326/acc25d}
}

@article{Chweeho2025NF,
  author = {Heo, C. and Park, S.J. and Choi, G. and Kim, J. and Kim, E.-J. and Jeon, Y.M. and Choi, M.J. and Han, H. and Sung, C. and Hahm, T.S. and Na, Y.-S. and the KSTAR Team},
  title = {Experimental identification of {I}-mode characteristics at the edge of {FIRE} mode in {KSTAR}},
  journal = {Nucl. Fusion},
  volume = {65},
  pages = {036003},
  year = {2025},
  doi = {10.1088/1741-4326/adacfc}
}

@article{Happel2019NME,
  author = {Happel, T. and Griener, M. and Silvagni, D. and Freethy, S.J. and Hennequin, P. and Janky, F. and Manz, P. and Prisiazhniuk, D. and Ryter, F. and Bernert, M. and Brida, D. and Eich, T. and Faitsch, M. and Gil, L. and Guimarais, L. and Merle, A. and Nille, D. and Pinz\'{o}n, J. and Sieglin, B. and Stroth, U. and Viezzer, E. and the ASDEX Upgrade Team and the EUROfusion MST1 Team},
  title = {Stationarity of {I}-mode operation and {I}-mode divertor heat fluxes on the {ASDEX Upgrade} tokamak},
  journal = {Nucl. Mater. Energy},
  volume = {18},
  pages = {159--165},
  year = {2019},
  doi = {10.1016/j.nme.2018.12.022}
}

@article{Wu2026NF,
  author = {Wu, C.B. and Ding, B.J. and Li, M.H. and Zou, X.L. and Zhong, X.M. and Liu, A.D. and Chu, Y.Q. and Wu, X.H. and Lin, S.Y. and Jin, Y.F. and the EAST Team},
  title = {Investigation of {LHCD} capabilities in {I}-mode plasma on {EAST} tokamak},
  journal = {Nucl. Fusion},
  volume = {66},
  pages = {026015},
  year = {2026},
  doi = {10.1088/1741-4326/ae2e15}
}

@article{Yu2025NF,
  author = {Yu, L. and Wang, L. and Zou, X.L. and Lin, X. and Meng, L.Y. and Yang, Q.Q. and Zhong, X.M. and Xu, L.Q. and Liu, A.D. and Liang, R.R. and Zhou, Z.Q. and Li, K.D. and Zang, Q. and Zhang, L. and Zhou, T.F. and Duan, Y.M. and Jin, Y.F. and Jia, T.Q. and Wang, M.R. and Liu, H.Q. and Zhang, B. and Xu, G.S. and Liu, Z.X. and Song, Y.T. and the EAST I-mode Working Group and EAST Team},
  title = {Towards detachment-compatible {I}-mode plasma on {EAST} tokamak},
  journal = {Nucl. Fusion},
  volume = {65},
  pages = {076005},
  year = {2025},
  doi = {10.1088/1741-4326/addc80}
}

@article{Wu2026PPCF,
  author = {Wu, D.G. and Yu, L. and Zou, X.L. and Meng, L.Y. and Wu, X.H. and Nan, K.B. and Jia, K. and Zang, Q. and Zhang, L. and Zhong, X.M. and the EAST Team},
  title = {Mechanism of core confinement improvement in {EAST} {I}-mode detachment experiments with neon seeding},
  journal = {Plasma Phys. Control. Fusion},
  volume = {68},
  pages = {045030},
  year = {2026},
  doi = {10.1088/1361-6587/ae59fa}
}

@article{BGWang2023NME,
  author = {Wang, Baoguo and Zhu, Dahuan and Ding, Rui and Gao, Binfu and Yan, Rong and Li, Changjun and Xuan, Chuannan and Yu, Baixue and Chen, Junling},
  title = {Observations on arcing on the metal plasma-facing components in {EAST}},
  journal = {Nucl. Mater. Energy},
  volume = {34},
  pages = {101318},
  year = {2023},
  doi = {10.1016/j.nme.2022.101318}
}

@article{Liu2014RSI,
  author = {Liu, H.Q. and Jie, Y.X. and Ding, W.X. and Brower, D.L. and Zou, Z.Y. and Li, W.M. and Wang, Z.X. and Qian, J.P. and Yang, Y. and Zeng, L. and Lan, T. and Wei, X.C. and Li, G.S. and Hu, L.Q. and Wan, B.N.},
  title = {Faraday-effect polarimeter-interferometer system for current density measurement on {EAST}},
  journal = {Rev. Sci. Instrum.},
  volume = {85},
  pages = {11D405},
  year = {2014},
  doi = {10.1063/1.4889777}
}

@article{ShenJ2013FED,
  author = {Shen, J. and Jie, Y. and Liu, H. and Wei, X. and Wang, Z. and Gao, X.},
  title = {Improved density measurement by {FIR} laser interferometer on {EAST} tokamak},
  journal = {Fusion Eng. Des.},
  volume = {88},
  pages = {2830--2834},
  year = {2013},
  doi = {10.1016/j.fusengdes.2013.05.019}
}

@article{Loarte2025PPCF,
  author = {Loarte, A. and Pitts, R.A. and Wauters, T. and Hughes, J.W. and Kukushkin, A.S. and Bilato, R. and Meigs, A. and de Vries, P.C. and Polevoi, A.R. and Kaveeva, E. and Rozhansky, V. and Belov, A. and the ITPA Group},
  title = {The new {ITER} baseline, research plan and open {R\&D} issues},
  journal = {Plasma Phys. Control. Fusion},
  volume = {67},
  pages = {065023},
  year = {2025},
  doi = {10.1088/1361-6587/add9c9}
}

@article{Zuo2013JNM,
  author = {Zuo, G.Z. and Hu, J.S. and Li, J.G. and Sun, Z. and Mansfield, D.K. and Zakharov, L.E.},
  title = {Lithium coating for {H}-mode and high performance plasmas on {EAST} in {ASIPP}},
  journal = {J. Nucl. Mater.},
  volume = {438},
  pages = {S90--S95},
  year = {2013},
  doi = {10.1016/j.jnucmat.2013.01.014}
}

@article{Guan2025NF,
  author = {Guan, Y.H. and Zuo, G.Z. and Xu, W. and Yu, Y.W. and Sun, Z. and Wang, Z. and Ding, R. and Zhang, L. and Zhang, T. and Wu, Z.W. and Mao, S.T. and Zhao, H.L. and Jia, T.Q. and Puyang, S.A. and Wang, L. and Wauters, T. and Gong, X.Z. and Hu, J.S.},
  title = {First result of boronization assisted by the {ICWC} on {EAST} with full metal wall},
  journal = {Nucl. Fusion},
  volume = {65},
  pages = {096020},
  year = {2025},
  doi = {10.1088/1741-4326/adf75c}
}

@article{Liu2025NF,
  author = {Liu, Yuming and Yan, Rong and Mu, Lei and Xuan, Chuannan and Wang, Baoguo and Sheng, Yuqi and Lin, Zichao and Zhou, Ziqiang and Zhu, Dahuan and Litnovsky, Andrey and Chen, Junling},
  title = {Studies of the boron erosion and deposition in shadowed areas in {EAST}},
  journal = {Nucl. Fusion},
  volume = {65},
  pages = {096024},
  year = {2025},
  doi = {10.1088/1741-4326/adf8ff}
}

@article{Cheng2025JINST,
  author = {Cheng, Yunxin and Zhang, Ling and Hu, Ailan and Morita, Shigeru and Zhou, Chengxi and Chen, Jihui and Zhang, Wenmin and Zhang, Fengling and Ma, Jiuyang and Liu, Haiqing},
  title = {Extreme ultraviolet spectrometers for fast observation of high-{Z} impurity line emissions and their radial profiles in the {Experimental Advanced Superconducting Tokamak}},
  journal = {JINST},
  volume = {20},
  pages = {C04016},
  year = {2025},
  doi = {10.1088/1748-0221/20/04/C04016}
}

@article{ZhangJ2023JINST,
  author = {Zhang, J. and Liu, A.D. and Zhou, C. and Zhuang, G. and Ding, W.X. and Hu, G.H. and Xu, G.S. and Ding, B.J. and Li, M.H. and Zou, X.L. and Zhang, S.B. and Hu, L.Q. and Feng, X. and Wang, S.F. and Wang, M.Y. and Li, H. and Lan, T. and Mao, W.Z. and Liu, Z.X. and Xie, J.L. and Liu, W.D.},
  title = {Determination of the radial position of zero density for profile reflectometry on experimental advanced superconducting tokamak},
  journal = {J. Instrum.},
  volume = {18},
  pages = {T01001},
  year = {2023},
  doi = {10.1088/1748-0221/18/01/T01001}
}

@article{Battaglia2013NF,
  author = {Battaglia, D.J. and Chang, C.S. and Kaye, S.M. and Kim, K. and Ku, S. and Maingi, R. and Bell, R.E. and Diallo, A. and Gerhardt, S. and LeBlanc, B.P.},
  title = {Dependence of the {L}--{H} transition on {X}-point geometry and divertor recycling on {NSTX}},
  journal = {Nucl. Fusion},
  volume = {53},
  pages = {113032},
  year = {2013},
  doi = {10.1088/0029-5515/53/11/113032}
}

@article{Zuo2012PPCF,
  author = {Zuo, G.Z. and Hu, J.S. and Zhen, S. and Li, J.G. and Mansfield, D.K. and Cao, B. and Wu, J.H. and Zakharov, L.E. and the EAST Team},
  title = {Comparison of various wall conditionings on the reduction of {H} content and particle recycling in {EAST}},
  journal = {Plasma Phys. Control. Fusion},
  volume = {54},
  pages = {015014},
  year = {2012},
  doi = {10.1088/0741-3335/54/1/015014}
}

@article{ITER1999NF,
  author = {{ITER Physics Expert Group on Confinement and Transport} and {ITER Physics Expert Group on Confinement Modelling and Database} and {ITER Physics Basis Editors}},
  title = {Chapter 2: Plasma confinement and transport},
  journal = {Nucl. Fusion},
  volume = {39},
  number = {12},
  pages = {2175--2249},
  year = {1999},
  doi = {10.1088/0029-5515/39/12/302}
}

@article{LiuY2019EPJ,
  author = {Liu, Yong and Li, E.Z. and Zhang, S.B. and Liu, H.Q. and Wang, S.X. and Hu, L.Q. and the EAST Team},
  title = {Overview of the {ECE} measurements on {EAST}},
  journal = {EPJ Web Conf.},
  volume = {203},
  pages = {03008},
  year = {2019},
  doi = {10.1051/epjconf/201920303008}
}

@article{Martin2008JPCS,
  author = {Martin, Y.R. and Takizuka, T. and the ITPA CDBM H-mode Threshold Database Working Group},
  title = {Power requirement for accessing the {H}-mode in {ITER}},
  journal = {J. Phys.: Conf. Ser.},
  volume = {123},
  pages = {012033},
  year = {2008},
  doi = {10.1088/1742-6596/123/1/012033}
}

@article{Shi2025PPCF,
  author = {Shi, W.X. and Zhou, C. and Liu, A.D. and Zhuang, G. and Zhang, S.B. and Li, X. and Wang, S.X. and Liu, H.Q. and Zhang, J. and Zhong, X.M.},
  title = {Measurement of multi-scale turbulence via {E}-band tunable ten-channel backscattering and one-channel forward-scattering integrated {Doppler} reflectometer on {EAST}},
  journal = {Plasma Phys. Control. Fusion},
  volume = {67},
  pages = {065014},
  year = {2025},
  doi = {10.1088/1361-6587/add9cc}
}

@article{XuZ2016RSI,
  author = {Xu, Z. and Wu, Z.W. and Gao, W. and Chen, Y.J. and Wu, C.R. and Zhang, L. and Huang, J. and Chang, J.F. and Yao, X.J. and Gao, W. and Zhang, P.F. and Jin, Z. and Hou, Y.M. and Guo, H.Y.},
  title = {Filterscope diagnostic system on the {Experimental Advanced Superconducting Tokamak} ({EAST})},
  journal = {Rev. Sci. Instrum.},
  volume = {87},
  pages = {11D429},
  year = {2016},
  doi = {10.1063/1.4961294}
}

@article{Conway2004PPCF,
  author = {Conway, G.D. and Schirmer, J. and Klenge, S. and Suttrop, W. and Holzhauer, E. and the ASDEX Upgrade Team},
  title = {Plasma rotation profile measurements using {Doppler} reflectometry},
  journal = {Plasma Phys. Control. Fusion},
  volume = {46},
  pages = {951--970},
  year = {2004},
  doi = {10.1088/0741-3335/46/6/003}
}

@article{Hirsch2001PPCF,
  author = {Hirsch, M. and Holzhauer, E. and Baldzuhn, J. and Kurzan, B. and Scott, B.},
  title = {{Doppler} reflectometry for the investigation of propagating density perturbations},
  journal = {Plasma Phys. Control. Fusion},
  volume = {43},
  pages = {1641--1660},
  year = {2001},
  doi = {10.1088/0741-3335/43/12/302}
}

@article{Viezzer2013NF,
  author = {Viezzer, E. and P\"{u}tterich, T. and Conway, G.D. and Dux, R. and Happel, T. and Fuchs, J.C. and McDermott, R.M. and Ryter, F. and Sieglin, B. and Suttrop, W. and Willensdorfer, M. and Wolfrum, E. and the ASDEX Upgrade Team},
  title = {High-accuracy characterization of the edge radial electric field at {ASDEX Upgrade}},
  journal = {Nucl. Fusion},
  volume = {53},
  pages = {053005},
  year = {2013},
  doi = {10.1088/0029-5515/53/5/053005}
}

@techreport{Zou1999TR,
  author = {Zou, X.L.},
  title = {Poloidal rotation measurement in {Tore Supra} by reflectometry},
  institution = {Association Euratom-CEA},
  year = {1999},
  type = {Technical Report}
}

@article{ZhangB2021NF,
  author = {Zhang, B. and Gong, X. and Qian, J. and Ding, R. and Huang, J. and Zou, X.L. and Liu, A.D. and Zhong, X.M. and Zhou, C. and Zhang, J.Y. and Jia, T.Q. and Liang, R.R. and Gao, W. and Zhong, G.Q. and Zeng, L. and Zhang, T. and Liu, H.Q. and Zang, Q. and Duan, Y.M. and Xu, L.Q. and Zhou, T.F. and Li, E.Z. and Li, M.H. and Xu, H.D. and Ding, B.J. and Song, Y.T. and Zhang, X.J. and Qin, C.M. and Wang, X.J. and Lyu, B. and Wang, L. and Zhang, L.},
  title = {{I}-mode operation in helium plasma with pure radio frequency wave heating and {ITER}-like tungsten divertor on {EAST}},
  journal = {Nucl. Fusion},
  volume = {61},
  pages = {116023},
  year = {2021},
  doi = {10.1088/1741-4326/ac207d}
}

@article{Song2023SciAdv,
  author = {Song, Y.T. and Zou, X.L. and Gong, X.Z. and B\'{e}coulet, A. and Buttery, R. and Bonoli, P. and Hoang, T. and Maingi, R. and Qian, J.P. and Zhong, X.M. and Liu, A.D. and Li, E.Z. and Ding, R. and Huang, J. and Zang, Q. and Liu, H.Q. and Wang, L. and Zhang, L. and Li, G.Q. and Sun, Y.W. and Garofalo, A. and Osborne, T. and Leonard, T. and Baek, S.G. and Wallace, G. and Xu, L.Q. and Zhang, B. and Wang, S.X. and Chu, Y.Q. and Zhang, T. and Duan, Y.M. and Lian, H. and Zhang, X.X. and Jin, Y.F. and Zeng, L. and Lyu, B. and Xiao, B.J. and Huang, Y. and Wang, Y. and Shen, B. and Xiang, N. and Wu, Y. and Wu, J.F. and Wang, X.J. and Ding, B.J. and Li, M.H. and Zhang, X.J. and Qin, C.M. and Xi, W.B. and Zhang, J. and Huang, L.S. and Yao, D.M. and Hu, Y.L. and Zuo, G.Z. and Yuan, Q.P. and Zhou, Z.W. and Wang, M. and Xu, H.D. and Xie, Y.H. and Wang, Z.C. and Chen, J.L. and Xu, G.S. and Hu, J.S. and Lu, K. and Liu, F.K. and Wu, X.C. and Wan, B.N. and Li, J.G.},
  title = {Realization of thousand-second improved confinement plasma with {Super} {I}-mode in {Tokamak} {EAST}},
  journal = {Sci. Adv.},
  volume = {9},
  pages = {eabq5273},
  year = {2023},
  doi = {10.1126/sciadv.abq5273}
}

\end{document}